# Line lists for the A$^2\Pi$-X$^2\Sigma^+$ (red) and B$^2\Sigma^+$-X$^2\Sigma^+$ (violet) Systems of CN, $^{13}$C$^{14}$N, and $^{12}$C$^{15}$N, and Application to Astronomical Spectra

Short title: CN A-X, B-X Transitions in Stellar Spectra

Christopher Sneden[1,2], Sara Lucatello[3], Ram S. Ram[4], James S. A. Brooke[4] and Peter Bernath[4,5]

[1]Department of Astronomy, University of Texas at Austin, Austin, TX 78712, USA; chris@verdi.as.utexas.edu

[2]Department of Astronomy and Space Sciences, Ege University, 35100 Bornova, İzmir, Turkey

[3]INAF, Osservatorio Astronomico di Padova, Vicolo dell'Osservatorio 5, I-35122 Padova, Italy; sara.lucatello@oapd.inaf.it

[4]Department of Chemistry, University of York, Heslington, York, YO10 5DD, UK; rr662@york.ac.uk; jsabrooke@gmail.com

[5]Department of Chemistry and Biochemistry, Old Dominion University, Norfolk, VA 23529 USA; pbernath@odu.edu

ABSTRACT

New red and violet system line lists for the CN isotopologues $^{13}C^{14}N$ and $^{12}C^{15}N$ have been generated. These new transition data are combined with those previously derived for $^{12}C^{14}N$, and applied to the determination of CNO abundances in the solar photosphere and in four red giant stars: Arcturus, the bright very low-metallicity star HD 122563, and carbon-enhanced metal-poor stars HD 196944 and HD 201626. When lines of both red and violet system lines are detectable in a star, their derived N abundances are in good agreement. The mean N abundances determined in this work generally are also in accord with published values.

.

# 1. INTRODUCTION

The abundances of carbon, nitrogen, and oxygen in evolved stars can be used to constrain stellar structure and evolution theories. Several decades ago Iben (1974 and references therein) undertook a major effort to model the interiors of stars, and to predict their surface CNO and $^{12}C/^{13}C$ changes from convective mixing episodes as they transform themselves from main sequence stars to subgiants, red giants, and horizontal branch (clump) stars. High-resolution spectroscopic surveys to test these predictions in different Galactic populations soon followed (e.g., Lambert & Ries 1977, 1981, Kjærgaard et al. 1982, Carbon et al. 1982, Cottrell & Sneden 1986). In recent years, renewed observational attention has been paid to these light elements in stars that are members of clusters (e.g., Lardo et al. 2012, Mikolaitis et al. 2012, Carrera & Martínez-Vázquez 2013), and those that reside in the field (e.g., Tautvaišienè 1997, Afşar et al. 2012, Tautvaišienè et al. 2012).

Surveys of stellar C and N abundances mostly rely on CH and CN molecular bands, augmented occasionally by the $C_2$ Swan bands in solar-metallicity disk stars and NH in low metallicity stars. However, the wavelength and transition probability accuracy of individual molecular lines has always been suspect, with little confidence in the completeness of line lists. Recently, greater emphasis has been placed in improving fundamental laboratory data for transitions that are of interest to stellar spectroscopy, both atomic (e.g. Wood et al. 2014 and references therein) and diatomic molecular: MgH (GharibNezhad et al. 2013, Hinkle et al. 2013), CH (Masseron et al. 2014), NH (Brooke et al. 2014b), $C_2$ (Brooke et al. 2013, Ram et al. 2014) and CN (Brooke et al. 2014a). In this paper we provide extensive new data for CN, with line lists for the $A^2\Pi$-$X^2\Sigma^+$ red and $B^2\Sigma^+$-$X^2\Sigma^+$ violet systems of two isotopologues, $^{13}C^{14}N$ and $^{12}C^{15}N$.

Studies of the CN red and violet systems published in the last couple of decades include those of Jorgensen & Larsson (1990), Hill et al. (2002), Kurucz (2011)[1].  However, the Jorgensen & Larsson work emphasized calculations of large numbers of CN red system lines for cool-star opacity work rather than accurate line wavelengths.  Some of the Hill et al. line positions have been revised recently following the laboratory investigation of the CN red system by Ram et al. (2010).  The Kurucz CN linelists were originally meant to provide good estimates of CN line blanketing in general, and as such have some substantial wavelength uncertainties.  The Kurucz database is in the process of undergoing an update to CN and other molecular line data.  We believe that it is important to provide a single internally-consistent set of CN red and violet line positions, excitation energies, and transition probabilities of the three major CN isotopologues; this is the aim of the present work.

In §2 we describe the construction of line lists for the CN red and violet systems.  Synthetic spectrum computations with these data are discussed in §3. We then apply these syntheses to the spectra of the solar photosphere in §3.1, the bright K-giant Arcturus in §3.2, the well-studied very metal-poor giant HD 122563 in §3.3, and two carbon-enhanced metal-poor giants in §3.4.

## 2. GENERATION OF CN LINE LISTS

Line lists for CN were generated as described by Brooke et al. (2014a) for the main isotopologue $^{12}C^{14}N$.  We used without change the line list that was provided as an electronic supplement to that paper.  In brief, high level ab initio transition dipole moment functions were calculated for the $A^2\Pi$-$X^2\Sigma^+$ and $B^2\Sigma^+$-$X^2\Sigma^+$ electronic transitions, as well as the dipole moment function for the $X^2\Sigma^+$ ground state. These functions and RKR (Rydberg-Klein-Rees) internuclear potential functions were used as input to Le Roy's program LEVEL (Le Roy 2007) to compute vibrational transition dipole matrix elements for a suitable range of v and J values. The matrix elements were calculated to cover the previous experimental observations of Ram et al. (2006, 2010a): v=0–22 for the $B^2\Sigma^+$ state, v=0–15 for the $A^2\Pi$ state and v=0–15 for the $X^2\Sigma^+$ state. The experimental line positions, and spectroscopic constants including the transition dipole matrix elements, were used with the PGOPHER[2] program (Western 2010) to compute line positions and line strengths for all possible vibrational bands for the range of v's listed above for the $A^2\Pi$-$X^2\Sigma^+$ and $B^2\Sigma^+$-$X^2\Sigma^+$ transitions, and the rovibrational bands of the ground $X^2\Sigma^+$ state. For these calculations, the rotational dependence of the transition dipole matrix elements as given by LEVEL was taken into account (Herman-Wallis effect). Additionally, the matrix elements were transformed from Hund's case (b) from LEVEL to Hund's case (a) for input into PGOPHER (Brooke et al. 2014a).

---

[1] **http://kurucz.harvard.edu/linelists.html**

Line lists for $^{13}C^{14}N$ and $^{12}C^{15}N$ were computed in the same way as described by Brooke et al. (2014a) using identical dipole moment functions and RKR potential energy curves as were adopted for $^{12}C^{14}N$. Within the Born-Oppenheimer approximation, these quantities are the same for all three isotopologues. For $^{13}C^{14}N$, the observed line positions are taken from Ram et al. (2010b) and Ram & Bernath (2011, 2012), and cover v=0-15 for the $B^2\Sigma^+$ state, v=0-22 for the $A^2\Pi$ state and v=0-15 for the $X^2\Sigma^+$ state. The experimental data for $^{12}C^{15}N$ are those of Colin and Bernath (2012) for v=0-3 for the $B^2\Sigma^+$ state, v=0-4 for the $A^2\Pi$ state and v=0-5 for the $X^2\Sigma^+$ state. The final output of these calculations consists of line positions, Einstein *A*-values and *f*-values for all possible bands (101 observed) for $^{13}C^{14}N$ (Table 1, see the online line list header for a detailed description), and all possible bands (28 observed) for $^{12}C^{15}N$ (Table 2). These data files also contain predictions (without hyperfine structure) for pure rotational transitions and vibration-rotation transitions in the $X^2\Sigma^+$ state.

---

[2] http://pgopher.chm.bris.ac.uk/

The $^{13}C^{14}N$ and $^{12}C^{15}N$ transition data of Tables 1 and 2 are also available from the first author in compact form (wavelength in Å, species identification code, excitation energy in eV, transition probability as log *gf*) suitable for use with the current version of the synthetic spectrum code MOOG (Sneden 1973).

## 3. CNO IN THE SUN AND IN FOUR EVOLVED STARS

Our computations applied standard LTE one-dimensional plane-parallel modeling techniques to spectral features that can be routinely detected in the spectra of most G-K type stars: the $C_2$ Swan bands, the CN red and violet systems, and the [O I] lines. Much more physically-realistic modeling studies have been accomplished for the solar photosphere (as summarized in the review of Asplund et al. 2009), and similar investigations of other types of stars have appeared in the recent literature (e.g., Ramírez et al. 2011, Bergemann et al. 2012, Lind et al. 2012, Kučinskas et al. 2013). However, many abundance studies will continue to employ ordinary modeling assumptions (with non-LTE correction factors applied to the LTE abundances) for the foreseeable future, especially in large stellar surveys.

The four giant stars chosen for study here present various challenges to the extraction of reliable CNO abundances. In this paper we do not attempt to provide definitive abundance sets for these stars; rather we aim to show the utility of the new molecular line lists in increasing the reliability of CNO determinations in each of these stellar abundance mixes.

For our analyses we used the current version of the LTE stellar spectrum line analysis code MOOG[3] (Sneden 1973), model stellar atmospheres discussed below, and line lists of atomic and molecular transitions, to generate synthetic spectra in the wavelength regions of interest. The synthetic spectra were then smoothed to match the combined effects of spectrograph resolution and stellar macroturbulence in the observed spectra.

The generation of synthetic spectrum line lists was previously described in our studies of MgH (Hinkle et al. 2013) and $C_2$ (Ram et al. 2014). We began with the atomic line compendium of Kurucz (2011), and substituted/added the transition data developed by the Wisconsin atomic physics group for neutron-capture elements (Sneden et al. 2009, and references therein) and for Fe-group elements (Wood et al. 2014, and references therein). In order to best reproduce the observed solar spectrum, we made small adjustments to the transition probabilities for lines with no laboratory data. For CN we adopted $D_0 = 7.724 \pm 0.002$ eV, which we computed from data contained in version 1.110 of the Active Thermochemical Tables[4]. The CN partition function is not needed here, because it does not enter into the computation of line strengths by MOOG, which follows the formalism of Schadee (1964). That is, full molecular equilibrium calculations must be done for a given model atmosphere and HCNO abundance set, but then only the resultant free number densities of the neutral C and N species

---

(and their partition functions) are needed to compute the CN strengths.  In all cases of interest here, the neutral atom number densities dominate in their molecular equilibria, followed by the CO density.  The CN density is always a trace contributor to the equilibrium calculations.  So while the partition function of CO needs to be reasonably well-determined, the CN partition function is largely irrelevant.

In the next several sections we give details of the syntheses for each star.  Table 3 summarizes the CNO abundances in log $\varepsilon$, [X/H], and [X/Fe] forms, as well as our estimates of $^{12}C/^{13}C$ ratios.  $^{12}C^{15}N$ lines are too weak to be detected in any of our program stars; the CN strengths are too small and the $^{14}N/^{15}N$ ratios are too large.  Hedrosa et al. (2013) have derived some N isotopic ratios in a sample of carbon-rich asymptotic-branch giant stars.  They employed a preliminary version of the $^{12}C^{15}N$ line list presented here, and thus the line parameters for the 11 lines of this isotopologue given in their Table 1 are essentially identical to our values.  The reader is referred to Figure 1 of Hedrosa et al. for examples of detection of $^{12}C^{15}N$ lines in extremely CN-rich spectra.

The [X/H] values are computed with respect to the Asplund et al. (2009) recommended photospheric abundances.  To calculate the [X/Fe] values, we adopted [Fe/H] metallicities from various literature sources.  Since this is an exploratory study on the application of new molecular line lists to cool stellar spectra of contrasting abundance mixes, we have made no attempt to derive new [Fe/H] values with our spectra, line lists, and stellar model atmospheres (whose parameters have been adopted from previous studies in the literature).  Therefore the [X/Fe] values are given only to one significant digit in Table 3.  Future studies of these stars to determine better relative abundances will be welcome.  However, abundance ratios among the CNO group, e.g. [C/O], can be viewed with greater confidence. We do not attempt full uncertainty analysis, but comment that for these stars, the maximum uncertainty in matching synthetic spectra of single absorption features to their observed counterparts ranges from 0.05 to 0.10 dex. However, since there are many lines of CN, $C_2$, and CH, the total synthetic/observed spectral matches can usually be estimated to within 0.05 dex. The well-determined dissociation energy (±0.002 eV) does not impact a CN error budget. The uncertainties in CN transition probabilities are discussed at length by Brooke et al. (2014), and on average appear to be about 5% for strong lines, contributing about ±0.02 to an N abundance uncertainty. This small value that is attributable to the parameters of the CN lines will be dwarfed by other error sources in a cool-star CNO analysis, such as the uncertainties in the O and C abundance determinations.

## 3.1 The Solar Photosphere

Renewed attention to the CNO group in the solar photosphere has led to significant revisions in the abundances of all three elements. The comprehensive summary by Anders & Grevesse (1989) recommended values of log ε(C)[5] = 8.56, log ε(N) = 8.05, and log ε(O) = 8.93, which were adopted by stellar spectroscopists for more than a decade. Fundamental re-examination of these values began with Allende Prieto et al. (2001), who derived a much smaller photospheric oxygen abundance from the reliable 6300.30 Å [O I] line:  log ε(O) = 8.69 ± 0.05.  Their study featured in-depth "three-dimensional time-dependent hydrodynamical" solar atmosphere modeling, but much of the downward abundance revision came from renewed attention to the contamination of the [O I] feature by a Ni I line at 6300.34 Å. Further studies of the photospheric O abundance (e.g. Caffau et al. 2008, Scott et al. 2009) have reached qualitatively similar conclusions.  Additionally, studies of N and C features in the solar spectrum (see below) have resulted in more modest decreases in those elemental abundances.  The recent review by Asplund et al. (2009) recommends adoption of photospheric abundances of log ε(C) = 8.43, log ε(N) = 7.83, and log ε(O) = 8.69, all with suggested uncertainties of ±0.05.

---

[5] We adopt standard spectroscopic notation. For elements X and Y, the relative abundances are written $[X/Y] = \log_{10}(N_X/N_Y)_{star} - \log_{10}(N_X/N_Y)_{\odot}$. For element X, the "absolute" abundance is written $\log \varepsilon(X) = \log_{10}(N_X/N_H) + 12$. Metallicity is taken to be the [Fe/H] value throughout this paper.

We employed the Holweger & Müller (1974) solar photospheric model to maintain consistency with our previous solar abundance studies that employ new atomic line data (Wood et al. 2014 and references therein) and molecular line data (Hinkle et al. 2013, Ram et al. 2014).  We adopted a microturbulent velocity of $v_{micro}$ = 1.1 km s$^{-1}$, but most molecular features in the solar spectrum are relatively weak, thus not much affected by the microturbulent velocity choice.

In principle the CNO abundances must be determined iteratively, since C and O are coupled through formation of the CO molecule, and CN features are contaminants for the [O I] lines at 6300, 6363 Å.  However, molecular equilibrium computations suggest that, in line-forming regions ($\tau \sim 0.5$) of the solar photosphere, the CO partial pressure is small compared to those of the atomic constituents:  $p(CO) \approx 0.02p(C) \approx 0.01p(O)$.  Thus CO exerts negligible influence on the number densities of either atom.  Additionally, CN features are detectable in the 6300 Å region, but they are very weak.  To a good approximation the CNO abundances can be determined independently for the solar photosphere.

*Oxygen:* only the 6300 Å line was employed. This transition provides the primary O indicator for all stars discussed in this paper. Details of synthetic/observed spectrum matches of the [O I] feature have been discussed in other papers and will not be repeated here. We adopted transition probabilities for the [O I] and Ni I lines recommended by Caffau et al. (2008), and with them estimated log ε(O) = 8.69, identical to the value for this line derived by Scott et al. (2009) and very close to the value from multiple O abundance features recommended by Caffau et al.

*Carbon:* our primary C abundance indicator is the $C_2$ $d^3\Pi_g$–$a^3\Pi_u$ Swan system. It has three detectable vibrational band sequences: $\Delta v = +1$, culminating in the blueward-degraded (1–0) P-branch head at 4737 Å; $\Delta v = 0$, with the (0–0) head at 5165 Å, and $\Delta v = -1$, with the (0–1) head at 5635 Å. We synthesized the solar spectrum in 20-Å intervals near the (0–0) and (1–0) bandheads, using the line lists developed by Brooke et al. (2013) and Ram et al. (2014). For all of our solar computations we assumed the solar-system isotopic ratio $^{12}C/^{13}C = 90$ (Wielen & Wilson 1997). In panels (a) and (b) of Figure 1 we show synthetic and observed spectra for small portions of these two bandhead regions. The observed spectra are part of the solar flux atlas of Wallace et al. (2011). Clearly the individual $C_2$ lines are very weak, and only the pileup of lines at the bandhead has substantial absorption depth. The average synthesis/observation matches to these spectra yielded log ε(C) = 8.53 and 8.48, respectively. We also synthesized the prominent CH $A^2\Pi$–$X^2\Delta$ G-band in the 4270–4330 Å interval, using the Masseron et al. (2014) molecular data. We began with their original transition table quantities and translated them into a table of values suitable for input to the MOOG spectrum synthesis code in the manner described in §2. In panel (c) of Figure 1 we show a small section of the CH spectrum. The lines here are very strong, but they yield consistent abundances, the mean of which is log ε(C) = 8.48, in accord with the results from $C_2$.

Our final C abundance gives most weight to the $C_2$ 5165 Å bandhead and CH: log ε(C) = 8.48. This value is only 0.05 dex larger than that recommended by Asplund et al. (2009) and is consistent with that derived by Caffau et al. (2010): log ε(C) = 8.50 ± 0.06.

*Nitrogen:* with our new line data we synthesized selected CN-rich regions in the solar spectrum. The CN $B^2\Sigma^+$–$X^2\Sigma^+$ violet system has well-known Δv = 0 bands led by the (0–0) and (1–1) bandheads at 3883 and 3871 Å, and Δv = −1 led by the (0–1) head at 4215 Å. In panel (a) of Figure 2 we show a small portion of our synthetic spectrum match to the (0–0) and (1–1) bandheads. There is much contamination by atomic species in this crowded spectral region, but the CN features consistently suggest log ε(N) = 8.08. We also synthesized the (0–1) bandhead, but it is very weak and blended with a strong Sr II resonance line. We estimate from this band that log ε(N) = 7.93, but do not give this result much weight.

The CN $A^2\Pi$-$X^2\Sigma^+$ red system has detectable absorption in the solar spectrum over a broad spectral interval, from at least 5800 Å to beyond 2 μm. In panels (b) and (c) of Figure 2 we show just two examples. The 8000 Å spectral region has been observed in many cool giant stars because it is the easiest wavelength region from which to estimate $^{12}C/^{13}C$ ratios. The 10970 Å spectral region was chosen at random from among many longer wavelength intervals that show strong CN absorption. The derived abundances are log ε(N) = 8.13 and 8.03 from the CN lines shown in panels (b) and (c) of Figure 2, and other regions show little N abundance variation. From all of our syntheses we suggest a mean abundance of log ε(N) = 8.05.

Our solar N abundance is somewhat larger than the value of log ε(N) = 7.83 recommended by Asplund et al. (2009), and log ε(N) = 7.86 ± 0.06 derived by Caffau et al. (2009).  Both of those studies relied primarily on analyses of N I lines.  For the absolute solar N abundance it is preferable to analyze transitions from the neutral atom, which is the dominant species in the N equilibrium.  CN is a trace species in line-forming regions of the solar photosphere:  our molecular equilibrium computations show that $p(CN)/p(N)$ $\sim 10^{-4}$.  Additionally of course, any abundance derived from CN can be determined no better than the assumed C abundance.  Combined with our standard modeling assumptions, it is obvious that caution should be observed in interpreting our derived mean N abundance.  However, we have shown here that a variety of CN features over a very large wavelength range yield consistent abundances.  CN is the molecule that provides most of the N abundances for other stars reported in the literature.

## 3.2 Arcturus

Arcturus is a very bright mildly metal-poor ([Fe/H] ~ −0.5) red giant star with extensive spectral atlases (Griffin 1968, Hinkle et al. 2000, Hinkle & Wallace 2005[6]).  This star has been subjected to several comprehensive abundance analyses (e.g., Mäckle  et al. 1975, Peterson et al. 1993, Ramírez & Allende Prieto 2011).  Special attention has been paid to the CNO elements, beginning with the discovery of an anomalously low C isotopic ratio by Lambert & Dearborn (1972): $^{12}C/^{13}C = 6.7 \pm 1.5$.  This result has been revised slightly upward in subsequent studies, e.g., Lambert & Ries (1977; $^{12}C/^{13}C = 7.2$); Pavlenko (2008; $^{12}C/^{13}C = 8 \pm 0.5$); Abia et al. (2012; $^{12}C/^{13}C = 9 \pm 2$).

---

[6] Available at ftp://ftp.noao.edu/catalogs/arcturusatlas/

But the essential result of a very low isotopic ratio for Arcturus has remained unchanged for four decades. Recently, Abia et al. (2012) summarized investigations of the Arcturus spectrum with special emphasis on the CNO group, and determined new values for this star's C and O isotopic ratios. They found $^{12}C/^{13}C = 9 \pm 2$ and log ε(C) = 8.06 ± 0.09, assuming log ε(N) = 7.67 ± 0.13 from Ryde et al. (2009), and log ε(O) = 8.76 ± 0.17 from Ramírez & Allende Prieto (2011).

For our computations we adopted the Arcturus model atmosphere parameters derived by Ramírez & Allende Prieto (2011): $T_{eff}$ = 4286 K, log $g$ = 1.66, [Fe/H] = –0.52, $v_{micro}$ = 1.74 km s$^{-1}$ as well as their individual abundances for elements other than the CNO group. These parameters were employed to generate a model atmosphere from the Kurucz (2011)[7] grid, using software kindly provided by A. McWilliam and I. Ivans. Models for the other stars are also interpolated from the Kurucz grid. The observed spectrum is the NOAO Arcturus Spectral Atlas (Hinkle et al. 2000). To derive the abundances, we used the line lists described in §3 or developed lists for new spectral regions in the same manner. We initially assumed $^{12}C/^{13}C = 7$ in the synthetic spectrum calculations; further comments on the isotopic ratio are given below.

*Oxygen:* in panel (a) of Figure 3 we show synthetic and observed spectra of the [O I] 6300 Å line. Transitions of CN can be easily identified in this panel by their strength responses to changes in O abundance. Specifically, a smaller O abundance decreases the 6300.30 Å [O I] absorption, and also decreases the number of C atoms that are tied up in CO. For typical line forming atmospheric depths ($\tau \sim 0.5$) in Arcturus, using our final CNO abundance set, we estimate that $p(C)/p(CO) \sim 0.5$ and $p(O)/p(CO) \sim 5$. A decrease in O liberates C from CO, resulting in more CN formation. This leads to increases in the strengths of prominent nearby CN features seen at 6297.5, 6298.6, 6304.1 Å, and in the many weaker transitions of this molecule. All elements of the CNO group take part in producing this and other spectral regions of Arcturus. Therefore it is necessary to derive the CNO abundances iteratively until satisfactory observed/synthetic matches are achieved. Iteration is needed especially to match the 6300.30 Å [O I] transition, which is contaminated both with the Ni I 6300.34 Å line discussed in §3.1 and a CN line at 6300.27 Å that can be ignored in the solar photosphere but not in Arcturus. Our final O abundance is log ε(O) = 8.63, well within the uncertainty suggested by Abia et al. (2012). We also synthesized the weaker 6363.78 Å [O I] line and derive an O abundance in excellent agreement with that derived from the 6300 Å line. However, there is significant CN absorption contaminating this [O I] feature, and this result should be treated with much caution.

---

[7] **http://kurucz.harvard.edu/grids.html**

*Carbon:* we again used the $C_2$ (0–0) and (0–1) bandheads as primary C abundance indicators, and used the CH G-band to confirm this result. Observed and computed $C_2$ spectra are illustrated in Figure 3 of Ram et al. (2014). From both of these bandheads we derived log ε(C) = 8.02, in good agreement with Abia et al. (2012). The CH lines are very strong and contaminated with many other strong atomic lines. Nevertheless, we estimate log ε(C) ~ 8.0 from CH, in accord with the more trustworthy result from $C_2$.

*Nitrogen*: the CN red system produces detectable features in all spectral regions redward of 6000 Å in Arcturus. We surveyed the CN absorption in several large spectral intervals: 6000–6100, 6280–6380, 8000–8100, and 10900–11000 Å. A small portion of the last of these regions is shown in panel (b) of Figure 3. All CN red system transitions considered here yield very similar abundances, the mean of which is log ε(N) = 7.66, with band-to-band scatter approximately ±0.05 dex. This result is supported by syntheses of the 4215 Å violet CN (0–1) bandhead, which we illustrate in panel (c) of Figure 3. In this very complex region it is difficult to find any clean CN transitions, but from consideration of the total absorption we derive log ε(N) ~ 7.6, matching the more solid result from the CN red system, and in good agreement with the value recommended by Abia et al. (2012).

$^{12}C/^{13}C$: Transitions of $^{13}C^{14}N$ are easy to detect in all spectral regions with relatively strong CN lines, $\lambda \geq 8000$ Å. Three small wavelength intervals with prominent $^{13}CN$ features are displayed in Figure 4. Panel (a) of this figure includes the triplet of $^{13}CN$ lines that have been employed most often in C isotopic studies of red giant stars. The C, N, and O elemental abundances are those derived above (Table 3). We have not performed a detailed analysis of these features, but our syntheses always suggest that $^{12}C/^{13}C = 7\pm1$, in accord with the literature sources cited above.

### 3.3 HD 122563

HD 122563 is the only very metal-poor star in the Bright Star Catalog (Hoffleit & Jaschek 1991). As such, its spectrum has been scrutinized at high resolution many times, beginning with Wallerstein et al. (1963). Sneden (1973) reported the first C and N abundances, finding [C/Fe] ~ −0.5 and [N/Fe] ~ +1.2. Following the assertion by Conti et al. (1967) that [O/Fe] > 0 in very metal-poor stars, Lambert et al. (1974) detected the [O I] line in HD 122563, deriving [O/Fe] ≈ +0.6. Subsequently, Lambert & Sneden (1977) used CH G-band lines to deduce $^{12}C/^{13}C = 5 \pm 2$, near the CN-cycle equilibrium value of 3–4 (e.g. Caughlan 1965). The HD 122563 abundances have been confirmed by subsequent studies. The most recent complete CNO determination is that of Spite et al. (2005), who derive [Fe/H] = –2.82, [C/H] = –3.29, [N/H] = –2.12 from CN or –1.72 from NH, and [O/H] = –2.20. Their CNO abundances translate to log ε(C) = 5.55, log ε(N) = 5.80 or 6.20, and log ε(O) = 6.77. The very low $^{12}$C/$^{13}$C ratio in HD 122563 has also been confirmed in other investigations, e.g. Sneden et al. (1986), Spite et al. (2006). The phenomenon of anomalous CNO-group abundances and C isotopic ratios in low-metallicity evolved stars (especially globular cluster giants) is now solidly established.

For our analysis we used an interpolated model with $T_{eff}$ = 4600 K, log $g$ = 1.30, [Fe/H] = –2.70, $v_{micro}$ = 2.5 km s$^{-1}$, close to the mean of parameters for HD 122563 in papers gathered in the SAGA database (Suda et al. 2011)[8]. The extreme metal deficiency leads to very few CNO-containing spectroscopic features that are available for abundance analysis. The spectrum used for our study was obtained with the Tull Echelle Spectrograph (Tull et al. 1995) of McDonald Observatory's 2.7 m Smith Telescope. The spectrograph resolving power is $R$ = 60,000.

*Oxygen*: Our synthesis of the [O I] 6300 Å transition is displayed in panel (a) of Figure 5. The Ni I and CN contaminants are undetectably weak in HD 122563; only O contributes at 6300.3 Å. We derived log ε(O) = 6.73, recovering the results of previous studies. We also detected the very weak 6363 Å [O I] line. Its absorption at line center is only 1% of the continuum, but our syntheses yield an abundance that is in agreement with that determined from the 6300 Å line.

*Carbon*: The C$_2$ Swan bands are undetectably weak in this very low metallicity star; we must rely only on the CH G-band for the C abundance.

---

We employed the same CH line list that was used for the Sun and Arcturus syntheses. A portion of synthesis/observation matches near the 4300 Å bandhead for HD 122563 is shown in panel (b) of Figure 5. A single C abundance works very well throughout the 4225–4375 Å spectral range: log ε(C) = 5.21.

*Nitrogen*: The CN red system is undetectably weak in the HD 122563 spectrum, and the very weak CN violet system (0–1) bandhead is lost among strong atomic absorption lines. The sole opportunity for deriving an N abundance lies with the 3883 Å (0–0) and 3871 Å (1–1) bandheads. Our observed/synthetic spectrum match of (0–0) lines is shown in panel (c) of Figure 5. Only very close to the bandhead can the CN absorption be established reliably. The derived abundance, log ε(N) = 5.85, also provides a good match to the very weak (1–1) bandhead in this star.

$^{12}C/^{13}C$: Test syntheses of CH spectral regions away from the bandhead (near 4230 and 4370 Å), where the $^{13}$CH and are significantly offset in wavelength from the $^{12}$CH lines, confirm previous work cited above that $^{12}C/^{13}C \approx 5$. The CN bands are too weak to permit detection of the $^{13}C^{14}N$ features.

The final abundance set for HD 122563 given in Table 3 is typical for ordinary evolved low-metallicity stars in the Galactic halo. Large-sample studies typically suggest that <[O/Fe]> ≈ +0.4 to +0.8 (e.g., Gratton et al. 2000, Spite et al. 2005); our value of [O/Fe] ≈ +0.7 equals the mean of the Spite et al. sample. The low C and high N abundances are also in accord with other work. We derive [N/C] ≈ +1.2 for HD 122563, while Gratton et al. obtain <[N/C]> ≈ +1.0 for the upper red giant branch stars in their sample, and Spite et al. obtain <[N/C]> ≈ +1.4 for their chemically-mixed stars.

### 3.4 Two CEMP Stars

HD 196944 and HD 201626 are two well-studied CEMP stars. Their low metallicities and large [C/Fe] ratios lead to dominance by C-containing molecular transitions. These CEMP stars should present good tests of the new $C_2$ and CN line lists. Spectra were obtained with HIRES at Keck telescope, with a resolving power of 48,000 and 103,000 for, respectively, HD 196944 and HD 201626.

Začs et al. (1998) performed an extensive study of HD 196944, and recommended $T_{eff}$ = 5250 K, log g = 1.7, [Fe/H] = –2.45, and $v_{micro}$ = 1.9 km s$^{-1}$. This star was included also in the CEMP high-resolution abundance analyses of Van Eck et al. (2003) and Aoki et al. (2002). All three investigations agree that the neutron-capture elements are very overabundant, e.g. [La/Fe] ~ [Ba/Fe] ~ +1.0. In our work we adopt the model atmosphere parameters of Aoki et al.: $T_{eff}$ = 5250 K, log $g$ = 1.8, [Fe/H] = –2.25, and $v_{micro}$ = 1.7 km s$^{-1}$. Those authors derived [C/Fe] = +1.2 and [N/Fe] = +1.3.

Vanture (1992a,b,c) studied the CNO element group in HD 201626, suggesting atmospheric parameters $T_{eff}$ = 5190 K, log $g$ = 2.25, [Fe/H] = –2.1, and $v_{micro}$ = 2.0 km s$^{-1}$. Van Eck et al. (2003) adopted these model parameters. Both studies determined very large neutron-capture overabundances for this star, e.g. [La/Fe] ~ +1.8. However, the $V$–$K$ color for HD 201626 in the SIMBAD[9] stellar database is 2.42, indicative of a lower temperature than HD 196944, for which $V$–$K$ = 1.87. If both of these stars are unreddened, then the temperatures derived from the Ramírez & Meléndez (2005) color calibration suggests $T_{eff}$ ≈4700 K for HD 201626 and 5200 K for HD 196944. Using the

---

[9] http://simbad.u-strasbg.fr/simbad/

Schlegel et al. (1998) or Schlafly & Finkbeiner (2011) extinction maps[10] we derive total Galactic reddening estimates $E(B–V) \approx 0.15$ or $E(V–K) \approx 0.41$ for HD 201626; the numbers are 0.04 and 0.11 for HD 196944. Then if both stars are suffering the total Galactic sightline reddening, their temperatures are really $T_{eff} \approx 5050$ K and 5420 K, respectively. But these objects do not reside at the edge of the Galaxy, and in particular the parallax of HD 201626 suggests that it is roughly only 250 pc away. Therefore we adopted here a compromise $T_{eff} = 4800$ K for HD 201626, and used the Padova isochrones (Girardi et al. 2010) to estimate $\log g = 1.3$ for a low metallicity star of this temperature. The other atmospheric parameters were assumed to be approximately [Fe/H] = –1.9 and $v_{micro} = 2.0$ km s$^{-1}$. We verified that the adopted metallicity is appropriate for this star by synthesizing the 6210–6270 Å spectral region, which has many strong Fe-group transitions and relatively weak CN absorption. This star is in need of a new comprehensive atmospheric analysis in the future.

*Carbon*: CEMP stars usually have [C/Fe] >> 0, so their C abundances can be determined independently of the assumed O abundances. We determined these abundances from the (0–0) and (0–1) $C_2$ bands and the CH G-band. In Figure 6 we show observed and computed spectra of $C_2$ lines in the 5148–5168 Å region. Dominance of $C_2$ can be easily seen in the relative weakness of the Mg I, Fe I blend at 5167.4 Å, which overwhelms the $C_2$ absorption in K-giant stars with ordinary CNO abundances (see Figure 3 of Ram et al. 2014). The $C_2$ lines are much stronger in HD 201626 than in HD 196944, due to its 450 K lower $T_{eff}$ and its larger C abundance; $C_2$ absorption is a sensitive function of both of these parameters. All C-containing features in both stars yield internally consistent abundances, and we obtain $\log \varepsilon(C) =$

---

[10] **http://irsa.ipac.caltech.edu/cgi-bin/bgTools/nph-bgExec**

7.36 for HD 196944 and log $\varepsilon$(C) = 7.88 for HD 201626.

*Oxygen*: The 6300 Å [O I] transition is detected in both stars, but it is only about 3% deep in HD 196944. The line suffers some blending by a telluric $O_2$ feature. We were able to cancel the contaminant to first order, but we also compared our synthetic spectra to observed spectra with and without the $O_2$ feature, and derived a similar O abundance in both cases. The [O I] line is deeper (~8%) in HD 201626, and is free from neighboring $O_2$ contaminants. We derived log $\varepsilon$(O) = 7.03 for HD 196944 and 7.28 for HD 201626. These values, translated into [O/Fe] ratios (Table 3) are consistent with the mean [O/Fe] overabundances of halo red giant stars as discussed in §3.3. The large production of C by the progenitors of these stars was not accompanied by any enhancement of O.

*Nitrogen*: The CN violet 4215 Å bandhead provides the most reliable N abundance in both stars. Synthetic and observed spectra for this spectral region are shown in Figure 7. For HD 196944, the 3883 Å band provides corroborating evidence, but the CN red system lines are too weak on our spectra to be useful. For HD 201626, the 3883 Å band is extremely strong, and provides no useful abundance information. But in compensation, the CN red system is easily detected in this star. From the 4215 Å bandhead we derived log $\varepsilon$(N) = 6.90, while from CN lines scattered throughout the 6200–6400 Å region the N abundances ranged from 6.95 to 7.05. The final abundance in Table 3 reflects our greater confidence in the CN red transitions.

$^{12}C/^{13}C$: Many of the branches of CH, $C_2$, and CN band systems have significant isotopic wavelength shifts, and our CEMP stars have such high relative C abundances ([C/Fe] ≈ +1.35, Table 3) that $^{12}C/^{13}C$ may be estimated from many spectral features. Ram et al. (2014) suggested that the $C_2$ Swan (0–0) bandhead region could be used to estimate the isotopic ratios for CEMP

stars; see their Figure 1.  Inspection of the gaps between the main $^{12}C_2$ lines reveals the presence of $^{12}C^{13}C$ lines when $^{12}C/^{13}C$ is low.  The same effect can be seen here in Figure 6.  The $C_2$ absorption is much less in HD 196944 than in HD 201626, but the strength ratio of the main and minor absorptions is different between the two.  The CH features near 4230 and 4370 Å were used also for the CEMP stars.  All syntheses yielded same results of $^{12}C/^{13}C \approx 6$ for HD 196944 and 25 for HD 201626.

Our new results are in reasonable agreement with previous studies of HD 196944; Začs et al. (1998) derived [C/Fe] = +1.4, in good agreement with our value in Table 3.  Their O abundance is much larger than ours:  [O/Fe] = +1.1.  However, their value was derived from an LTE analysis of the high-excitation O I triplet.  Many studies have shown that severe departures from LTE attend the formation of the triplet lines in low metallicity stars:  Fabbian et al. (2009) suggest that calculated [O/Fe] LTE abundance ratios are too large by 0.5 to 0.9 dex.  If so, our much lower abundance deduced from the [O I] line is consistent with their result.

For HD 201626, our absolute CNO abundance scale is about 0.5 dex lower than that of Vanture (1992b).  However, his abundance ratios are in better agreement with ours. From his Table 3, $\varepsilon(C)/\varepsilon(O) = 4.0$  and $\varepsilon(C)/\varepsilon(N) = 5.6$, while the data from Table 3 yield  $\varepsilon(C)/\varepsilon(O) = 4.0$ and $\varepsilon(C)/\varepsilon(N) = 8.0$.  The difference in C/N ratio may simply be due to a difference in CN line lists, and is not large enough to pursue further.  Additionally, we concur with Vanture's (1992a) assessment that $^{12}C/^{13}C \geq 25$ for this star.

## 4. CONCLUSIONS

We have computed new line lists for the red $A^2\Pi$-$X^2\Sigma^+$ and violet $B^2\Sigma^+$-$X^2\Sigma^+$ band system of the CN isotopologues $^{13}C^{14}N$ and $^{12}C^{15}N$, to complement the line lists for the main $^{12}C^{14}N$ isotopologue that have been presented by Brooke et al. (2014a). These line lists are available in detail with this paper and in a form that can be obtained from the authors that can be readily used with stellar synthetic spectrum codes. We have explored the utility of these new CN data by deriving CNO abundances and $^{12}C/^{13}C$ isotopic ratios in the solar photosphere and in four red giant stars: Arcturus, a mildly metal-poor probable thick disk member; HD 122563, a very metal-poor halo star; and HD 196944 and HD 201626, two carbon-enhanced metal-poor (CEMP) halo stars. Individual CN transitions in the synthetic spectra that we have generated match well the observed spectra in wavelengths and absorption strengths. The CN red system features, which are spread over a very large range of the red and infrared spectral regions, yield consistent N abundances and $^{12}C/^{13}C$ ratios. Additionally, the N abundances derived from the red and violet systems are generally in agreement when such comparisons can be attempted.

## ACKNOWLEDGEMENTS

We thank Melike Afşar, Gamze Bocek Topcu, Marc Schaeuble, and the referee for helpful comments on the manuscript. Some of data presented herein were obtained at the W.M. Keck Observatory, which is operated as a scientific partnership among the California Institute of Technology, the University of California and the National Aeronautics and Space Administration. The Observatory was made possible by the generous financial support of the W.M. Keck Foundation. This research was supported by the Leverhulme Trust of the UK. Additional funding was provided by the NASA laboratory astrophysics program. This work was also supported in part by NSF Grant AST-1211585 to C.S., and S.L. acknowledges partial support from PRIN INAF 2011. Major parts of this research were undertaken while C.S. was on a University of Texas Faculty Research Assignment, in residence at the Department of Astronomy and Space Sciences of Ege University. Funding from the University of Texas and The Scientific and Technological Research Council of Turkey (TÜBITAK, project No. 112T929) are greatly appreciated. Some of the spectra used in this work were recorded at the National Solar Observatory at Kitt Peak, USA.

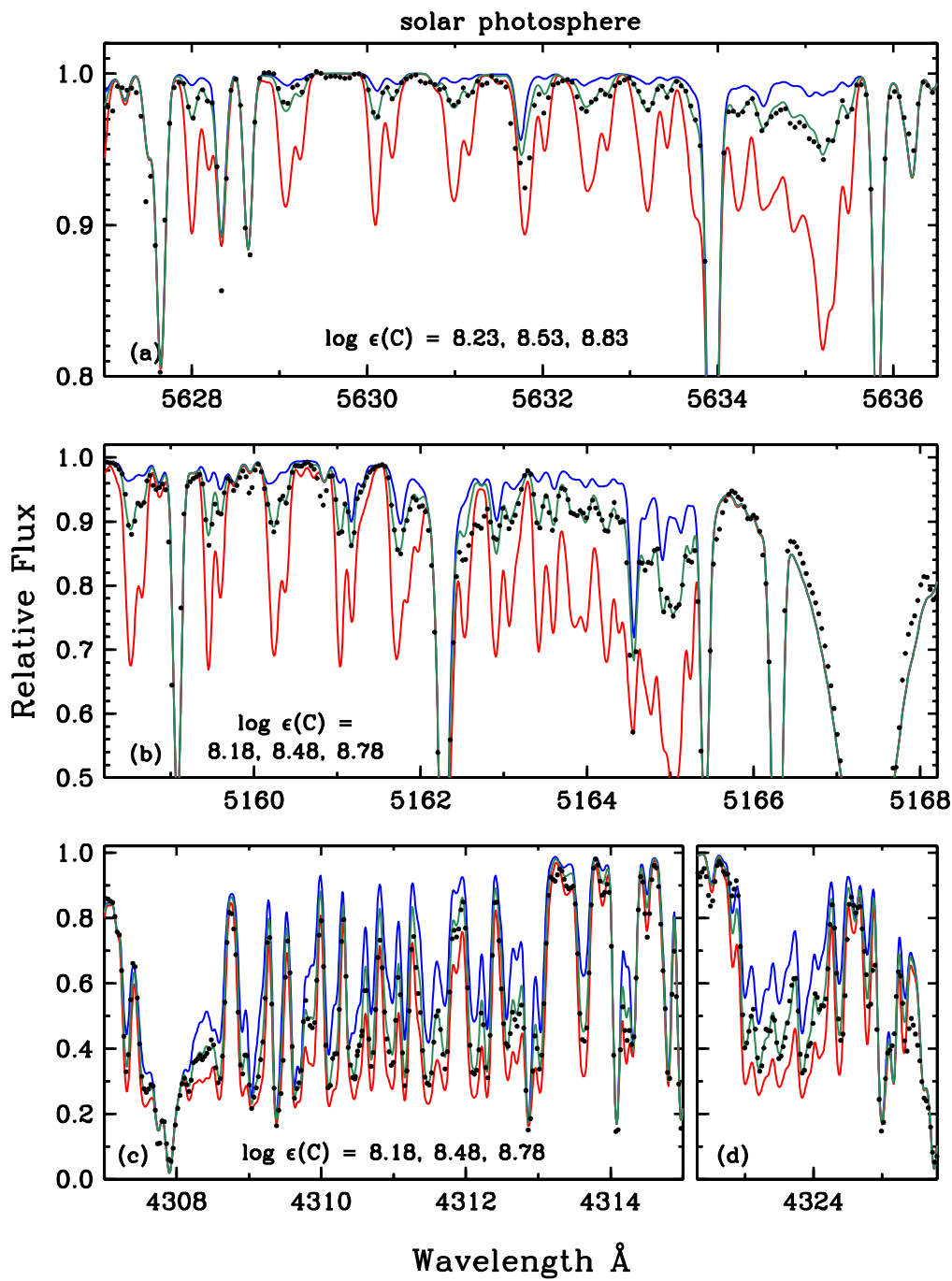

Figure 1: Synthetic spectra (colored lines) and observed solar photospheric spectra (black points) of three wavelength regions with $C_2$ (panels a, b) and CH (panels c, d) absorption features. In each panel, the smallest, best fit, and largest abundances are indicated with blue, dark green, and red lines, and their values are written in the panel legend.

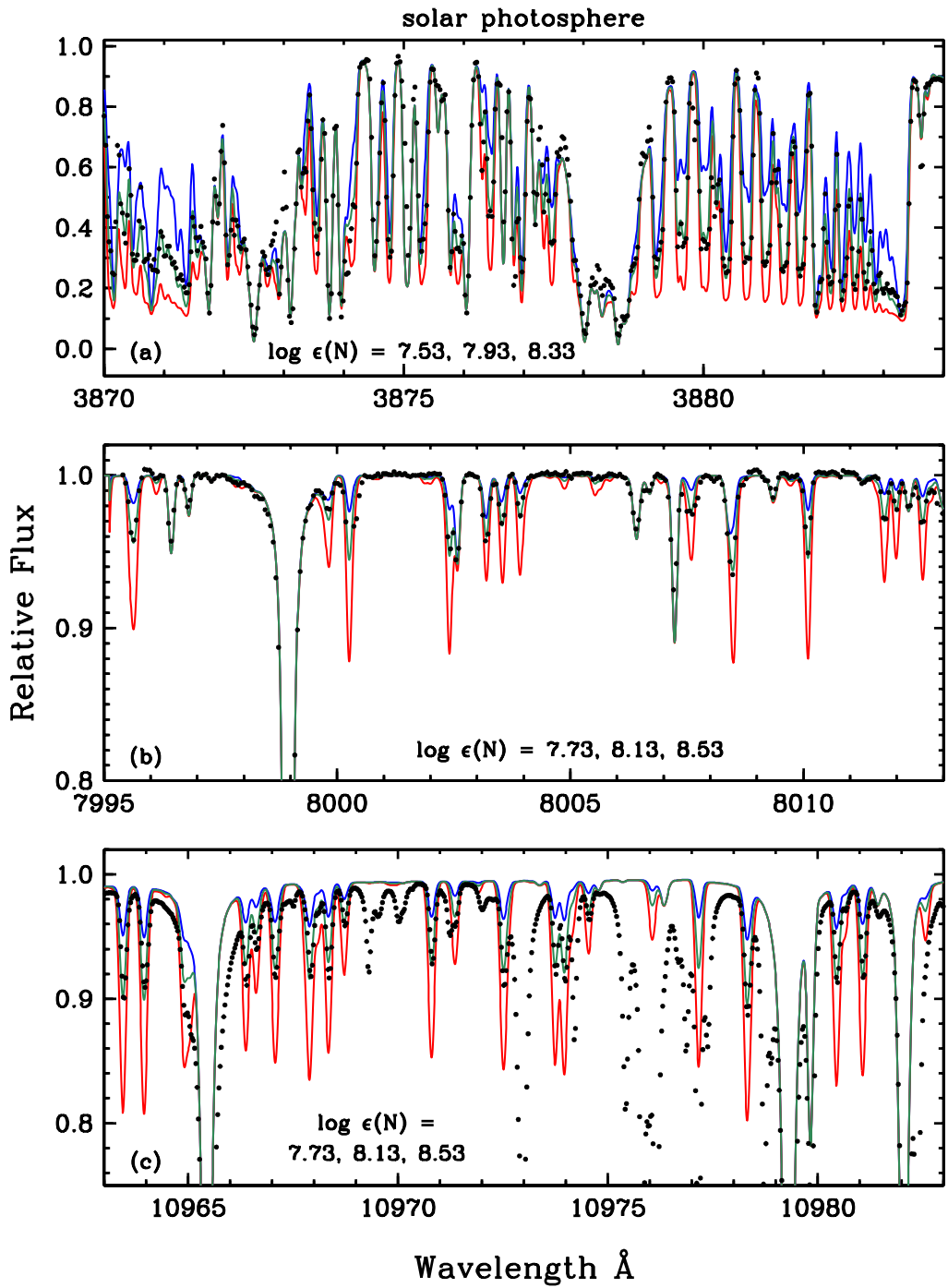

Figure 2: Synthetic and observed solar photospheric spectra of three wavelength regions with CN absorption features. Points and lines are as in Figure 1. The observed features in panel (c) not well matched by the syntheses are telluric in origin.

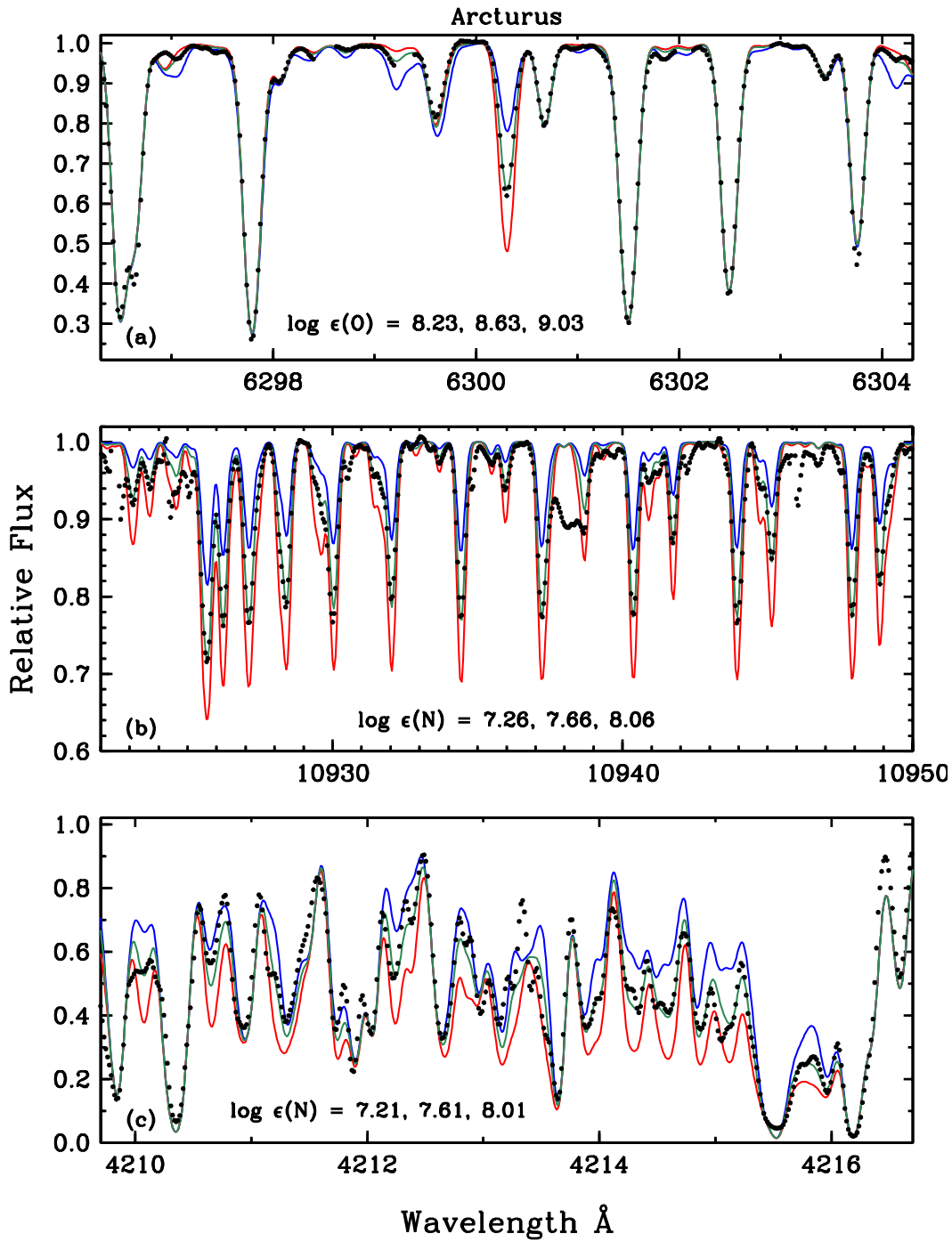

Figure 3: Synthetic and observed Arcturus spectra of the [O I] 6300 Å line (panel a), a portion of the CN red system (0–0) R-branch bandhead (panel b), and a portion of the CN violet system (0–1) bandhead (panel c). Points and lines are as in Figure 1.

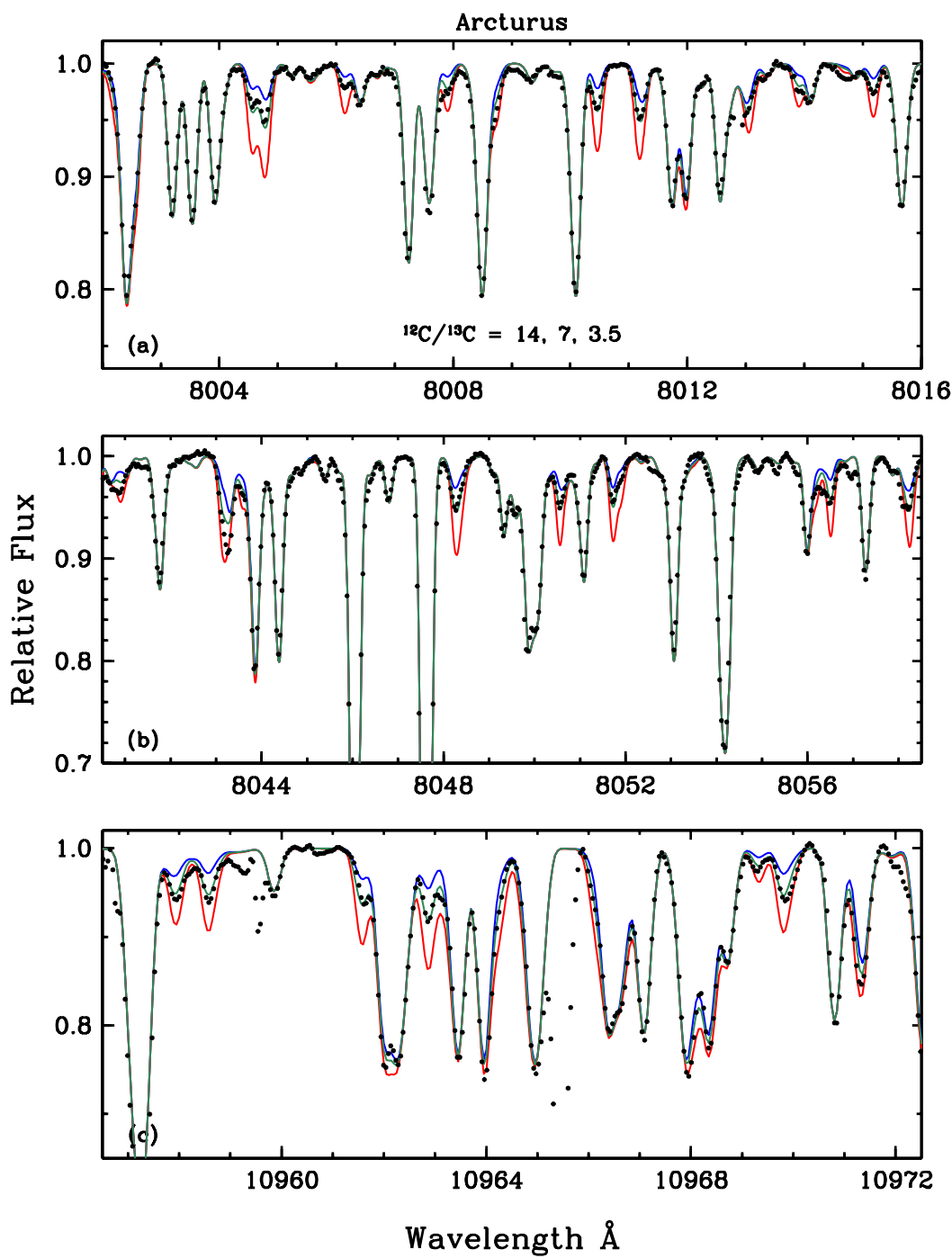

Figure 4: Synthetic and observed Arcturus spectra in three regions of the CN red system that have prominent $^{13}$CN absorption features. The assumed $^{12}$C/$^{13}$C ratios for the syntheses are as labeled in panel (a). Points and lines are as in Figure 1.

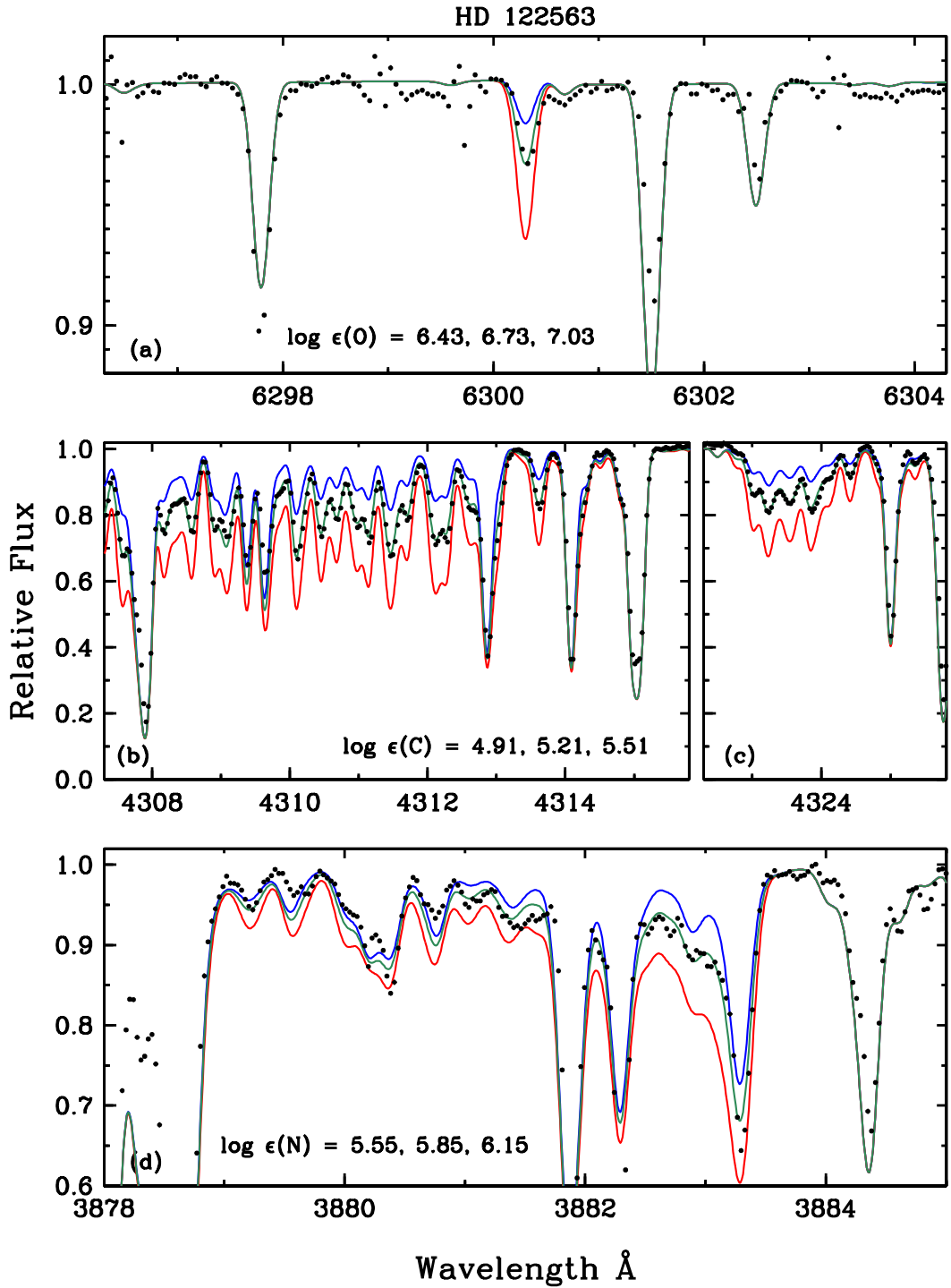

Figure 5: Synthetic and observed HD 122563 spectra of the [O I] 6300 Å line (panel a), portions of the CH G-band (panels b and c), and the CN violet system (0–0) bandhead (panel d). Points and lines are as in Figure 1.

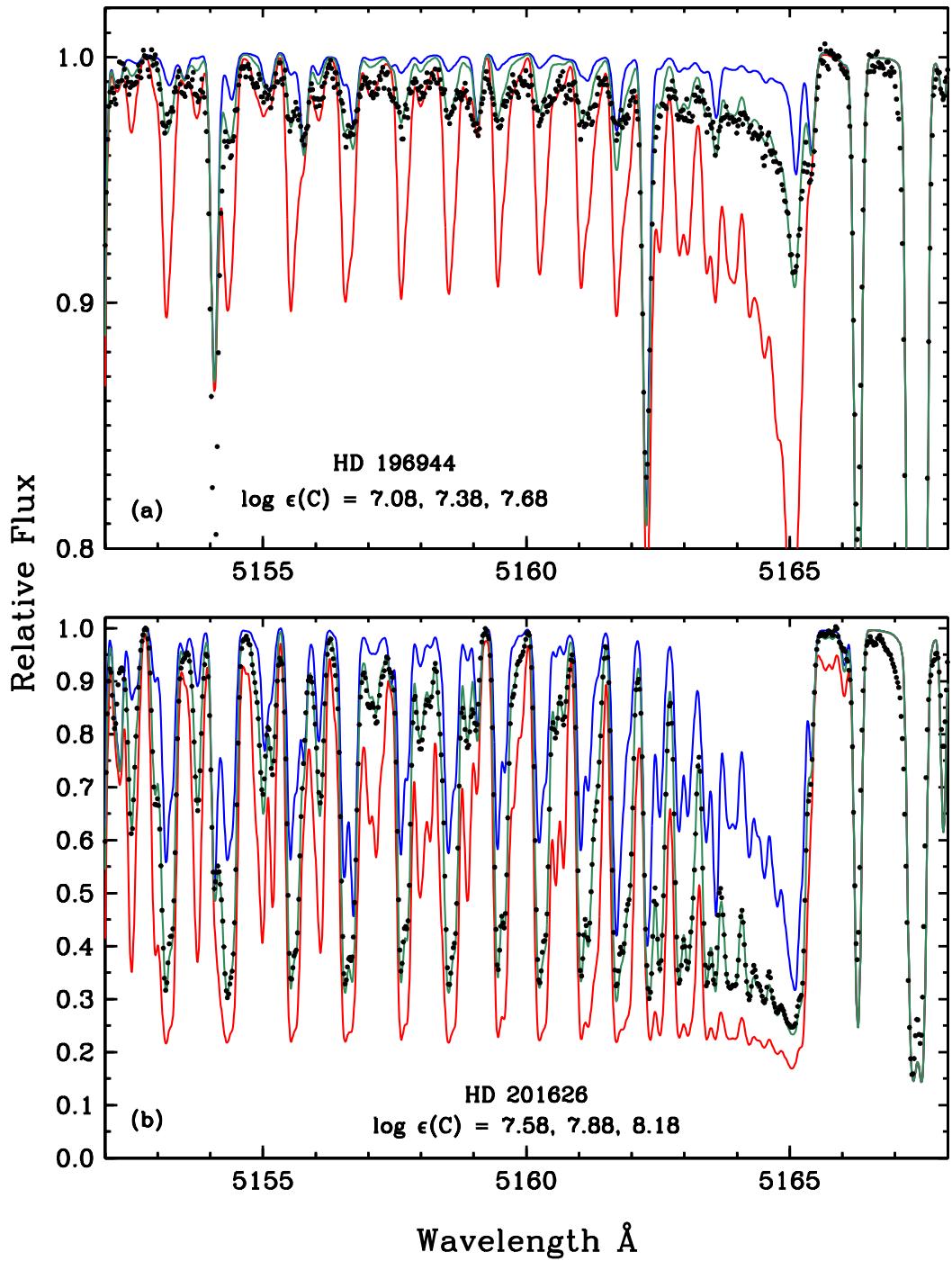

Figure 6: Synthetic and observed spectra of the 5155 Å C$_2$ (0–0) bandhead in two CEMP stars. Points and lines are as in Figure 1.

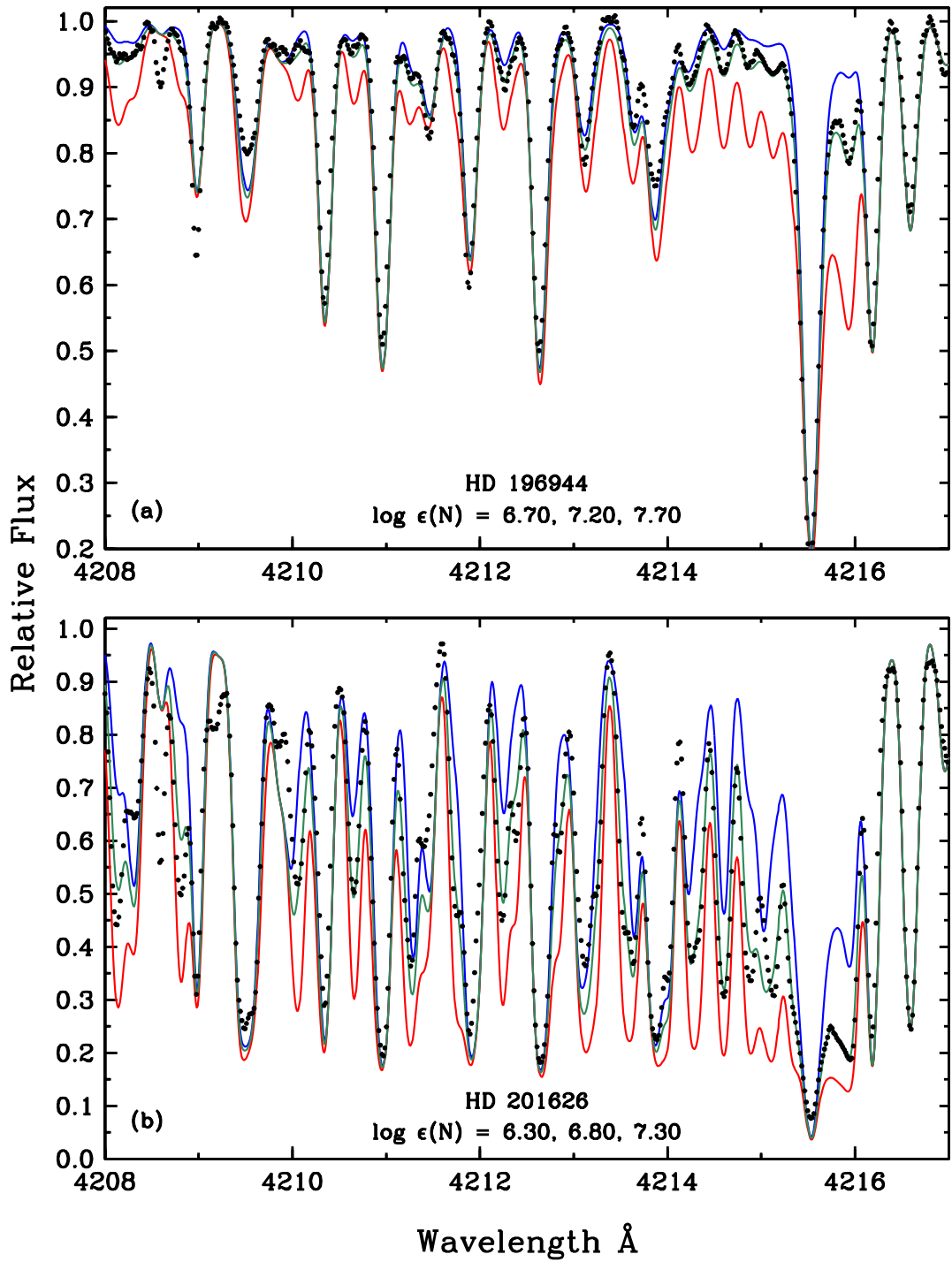

Figure 7: Synthetic and observed spectra of the 4215 Å CN violet system (0–1) bandhead in two CEMP stars. Points and lines are as in Figure 1.